\documentstyle[twocolumn,psfig,aps]{revtex}
\begin{document}

\def\gtwid{\mathrel{\raise.3ex\hbox{$>$\kern-.75em\lower1ex\hbox{$\sim$}}}}
\def\ltwid{\mathrel{\raise.3ex\hbox{$<$\kern-.75em\lower1ex\hbox{$\sim$}}}}
\def\ed{{\sl edited\ by\ }}
\def\cf{{\sl cf.~}}
\def\viz{{\sl viz.~}}
\def\etal{{\sl et.~al.}}
\def\tj{$t$-$J$~}
\twocolumn[\hsize\textwidth\columnwidth\hsize\csname
@twocolumnfalse\endcsname

\title{Evolution of the Spin Gap Upon Doping a 2-Leg Ladder}

\author{D.~Poilblanc$^a$\thanks{didier@irsamc2.ups-tlse.fr}, O.~Chiappa$^a$ 
and J.~Riera$^b$\thanks{jose@jaguar.fceia.unr.edu.ar},}
\address{
$^a$Laboratoire de Physique Quantique \& UMR--CNRS 5626,
Universit\'e Paul Sabatier, F-31062 Toulouse, France\\
$^b$Instituto de F\'{\i}sica Rosario, Consejo Nacional de
Investigaciones 
Cient\'{\i}ficas y T\'ecnicas, y Departamento de F\'{\i}sica,
Universidad Nacional de Rosario, Avenida Pellegrini 250, 2000-Rosario,
Argentina
}
\author{S.R.~White$^c$\thanks{srwhite@uci.edu} and 
D.J.~Scalapino$^d$\thanks{djs@vulcan.physics.ucsb.edu}}
\address{$^c$Department of Physics and Astronomy, University of California,
Irvine, CA 92697 \\
$^d$Department of Physics, University of California, Santa Barbara CA
93106}
\date{\today}
\maketitle

\begin{abstract}

The evolution of the spin gap of a 2-leg ladder upon doping depends upon
the nature of the lowest triplet excitations in a ladder with two holes.
Here we study this evolution using various numerical techniques
for a $t$-$t^\prime$-$J$
ladder as the next-near-neighbor hopping $t^\prime$ is varied.  We find
that depending on the value of $t^\prime$, the spin gap can evolve
continuously or discontinuously and the lowest triplet state can correspond
to a magnon, a bound magnon-hole-pair, or two separate quasi-particles.
Previous experimental results on the superconducting 
two-leg ladder Sr$_{14-x}$Ca$_x$Cu$_{24}$O$_{41}$
are discussed.

\smallskip
\noindent PACS: 71.27.+a, 75.50.Ee, 71.10.-w, 75.40.Mg
\end{abstract}

\vskip2pc]


Studies of strongly-correlated electrons confined to two-leg ladders and
described by \tj and Hubbard models have provided important insights
into the high $T_c$ cuprate puzzle. 
These models are known to exhibit a
gaped spin liquid state at half-filling and upon doping
to evolve into a Luther-Emery
state characterized by $d_{x^2-y^2}$-like pairing and $4k_F$ CDW
correlations \cite{DR96,HPNSH95}. 
A key feature of the Luther-Emery state is the
existence of a gap $\Delta_S$ in the excitation energy of the
spin degrees of freedom.  At
half-filling, the spin gap $\Delta_S$ is set by the 
${\mathbf K}=(\pi,\pi)$ magnon
excitation energy $\Delta_{M}$ which is of order $J/2$ for an isotropic
Heisenberg ladder with a near neighbor exchange interaction $J$.  However,
as discussed by Tsunetsugu \etal \cite{TTR94}, 
there can be a discontinuous evolution of
the spin gap upon doping.  In particular, they note that a pair can be
dissociated into two charge $|e|$ and spin $S=1/2$ quasi-particles, and the
low-energy continuum for such scattering is set by the pair-binding energy
$\Delta_P$. Then,
if the pair-binding energy is less than the half-filled spin gap ($\Delta_P
< \Delta_M$), there
will be a discontinuous decrease in the spin gap upon doping to a value
equal to the pair-binding energy $\Delta_{P}$.  Thus, while there is still
an $S=1$, ${\mathbf K}=(\pi,\pi)$ 
magnon excitation with energy $\Delta_M$ in the
infinitesimally doped ladder, if $\Delta_M > \Delta_P$ a lower energy $S=1$
state exists in which a pair is dissociated into two quasi-particles.

There is a low energy continuum of excited states corresponding to two
quasi-particles, each in an even parity $k_y=0$ state, which have
a total momentum $(k_x,k_y)=(0,0)$. Here $k_y=0$
for a bonding and $\pi$ for an anti-bonding quasi-particle respectively.
The singlet and triplet continua start at the same energy $\Delta_P$.
In addition to these scattering states, there can also be a bound $S=1$
state in which a bonding and an anti-bonding quasi-particle with momentum
$k_y=\pi$ hybridize with a magnon excitation of the spin background
\cite{TTR96,LBF98}. If there is such a 
bound magnon-pair with energy $\Delta_{MP}$, then it will set
the spin gap in the doped ladder provided
$\Delta_{MP}<\Delta_M$ \cite{note1}.  
Such a scenario occurs e.g. in ladders with anisotropic rung ($J_\perp$) and
leg ($J_\parallel$) couplings in the 
range $0.4 \ltwid J_\perp/J_\parallel \ltwid 1.4$ \cite{RPD99}. 
Here we combine
exact diagonalization (ED) \cite{HP96}
and density-matrix-renormalization-group (DMRG) techniques \cite{Whi93} to
investigate the evolution of the spin gap when one pair of holes is added
to a $t$-$t^\prime$-$J$ ladder. The next-near-neighbor
one-electron hopping $t^\prime$ provides a useful tuning parameter to study the
interplay of the magnon gap $\Delta_{M}$, the pair
dissociation gap $\Delta_{P}$, and the bound magnon-pair spin gap
$\Delta_{MP}$ in setting the spin gap $\Delta_S$ of the lightly doped
ladder.

The Hamiltonian for the $t$-$t^\prime$-$J$ ladder is
\begin{eqnarray}
H&=&J \sum_{i,\lambda} (\vec{\mathbf S}_{i,\lambda}\cdot
\vec{\mathbf S}_{i+1,\lambda}-\frac{1}{4}n_{i,\lambda}n_{i+1,\lambda}) \\
&+&J\sum_{i} (\vec{\mathbf S}_{i,1}\cdot \vec{\mathbf S}_{i,2}
-\frac{1}{4}n_{i,1}n_{i,2})\nonumber \\
+&t& \sum_{i,\lambda,s}
(c_{i,\lambda,s}^\dagger c_{i+1,\lambda,s}
+ h.c.)+t \sum_{i,s} (c_{i,1,s}^\dagger
c_{i,2,s}+h.c.) \nonumber \\
&+&t' \sum_{i,s} (c_{i,1,s}^\dagger
c_{i+1,2,s}+c_{i,2,s}^\dagger
c_{i+1,1,s}+h.c.) \nonumber \, ,
\label{one}
\end{eqnarray}
Here $c^\dagger_{i,\lambda, s}$ creates an electron of spin $s$ on site $i$
of leg $\lambda=1$ or 2, $\vec {\mathbf S}_{i,\lambda} = 
 (c^\dagger_{i,\lambda,s}
\ \vec \sigma_{ss^\prime} c_{i,\lambda,s^\prime})/2$ 
and $n_{i,\lambda}= \Sigma_s c^\dagger_{i,\lambda,s}
c_{i,\lambda,s}$.  We have taken both the near-neighbor leg and rung
one-electron hopping matrix elements equal to $t$ and the diagonal
next-near-neighbor term equal to $t^\prime$.  The exchange interaction $J$
is taken as isotropic between near-neighbor leg and rung sites and
throughout this $J/t=0.5$.

We begin with our conclusions shown in Fig.~1(a) where we have plotted the
excitation energies $\Delta E$ of various triplet states versus $t^\prime$. 
The spin gap $\Delta_S$
of the two leg $t$-$t^\prime$-$J$ ladder doped with two holes is defined as
the difference between the ground state energies of the system with two
holes and $S=1$ and $S=0$ respectively.
\begin{equation}
\Delta_S = E_0 \left(n_h=2, S=1\right) - E_0 \left(n_h=2, S=0\right)\ .
\label{two}
\end{equation}
The stars in Fig.~1(a) show $\Delta_S$ versus $t^\prime/t$ obtained
from DMRG results on $2\times L$ ladders with $L=32$.
The dashed line is the DMRG
result for the magnon excitation of the undoped ladder obtained from
\begin{equation}
\Delta_{M} = E_0 \left(n_h=0, S=1\right) - E_0 \left(n_h=0, S=0\right)
\ .
\label{three}
\end{equation}
That this difference in ground state energies corresponds to the
$(\pi,\pi)$ magnon is known from ED
calculations in which the momentum
of the excitation is specified.  The open diamonds show the triplet
excitation energy in the ${\mathbf K}=(\pi,\pi)$ sector, obtained from
a finite size scaling analysis using ED.
Finally, the solid curve in Fig.~1
corresponds to the pair-binding energy 
calculated with DMRG from
\begin{eqnarray}
\Delta_{P}&=& E_0 \left(n_h=2, S=0\right) + E_0 \left(n_h=0, S=0\right) 
\nonumber \\
&-& 2E_0 \left(n_h=1, S=\frac{1}{2}\right)
\label{four}
\end{eqnarray}
with $n_h$ the number of holes relative to the half-filled ladder
(in agreement with the ED results for $t'=0$ in Ref.~\cite{RPD99}).
As shown in Fig.~1(b), $\Delta_P$ sets the two quasi-particle continuum.
Here infinite size extrapolated ED results for the lowest energy excited
singlet and triplet states in the ${\mathbf K}=(0,0)$ sector are plotted as 
open symbols and the solid circles are DMRG data for the pair-binding energy
$\Delta_P$, Eq.~(\ref{four}). These energies are in good agreement,
consistent with a picture in which a pair dissociates into two
quasi-particles. 

As discussed below, we have used ED, in which the momentum
of the state can be specified, as well as DMRG calculations of the hole and
spin correlations in order to interpret the results shown in Fig.~1(a).   Here
we summarize what these show. Basically, there are three different regimes
set by $t^\prime/t$.  For $-0.5 < t^\prime/t \ltwid -0.2$, the discontinuous
drop in the spin gap with doping reflects the fact that the pair binding
energy $\Delta_{P}$ is less than the $(\pi,\pi)$ magnon energy $\Delta_M$
of the undoped ladder.  Thus, when the system is doped, a singlet pair can
dissociate into two separate quasi-particles with total spin $S=1$,
reducing the spin gap $\Delta_S$ from $\Delta_M$ to $\Delta_P$.
In this region, there is a bound magnon-hole pair with a minimum energy at
$(\pi,\pi)$ but its energy $\Delta_{MP}$ is larger than $\Delta_P$
so that the spin gap is set by $\Delta_P$.
In the region $-0.2 \leq t^\prime/t \leq 0.35$, the situation changes. The
pair binding energy $\Delta_P$ becomes greater than the energy to create a
bound magnon-hole pair $\Delta_{MP}$, but $\Delta_{MP}$ is less than the
energy to create a separate magnon $\Delta_M$. Thus, in  this parameter
region
the lowest energy triplet state of the 2-hole doped ladder has momentum
$(\pi,\pi)$ and corresponds to a bound magnon-hole pair so that $\Delta_S =
\Delta_{MP}$.
Finally, for $0.35 < t^\prime/t < 0.5$, the energy of the
triplet ${\mathbf K}=(\pi,\pi)$ excitation becomes equal to the $S=1$ 
magnon energy
of the undoped ladder.  Here DMRG calculations of the spin and charge
correlations show that the excitation corresponds to a magnon which is
uncorrelated with the bound singlet pair. Thus, in this region, there is no
discontinuity in the spin gap upon doping. 
\begin{figure}
\begin{center}
\psfig{figure=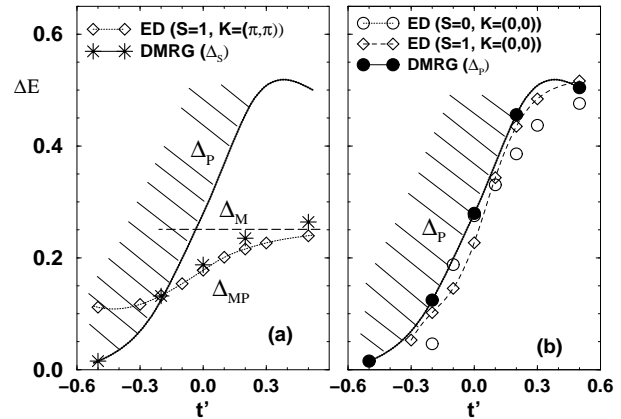,width=8truecm,angle=-90}
\end{center}
\vspace{-0.2truecm}
\caption{(a): Low energy triplet excitations of a 2-leg $t$-$J$
ladder versus $t^\prime$ for $J/t=0.5$. The shaded region corresponds to the
onset of the two quasi-particle continuum set by the pair-breaking energy
$\Delta_P$. The dashed line denotes the magnon energy $\Delta_M$ of the
undoped ladder. The open diamonds show ED results for the lowest triplet
excitation in the ${\mathbf K}=(\pi,\pi)$ sector and the stars denote the DMRG
spin gap $\Delta_S$ measured for a ladder with 2 holes. 
(b): the open symbols show ED results for the lowest energy singlet and
triplet excitations in the ${\mathbf K}=(0,0)$ sector.  The solid circles show
DMRG data for the pair-binding energy $\Delta_P$, Eq.~(\ref{four}). The
solid line through the DMRG points denotes the two quasi-particle continuum
as also shown in (a). 
}
\label{scaling}
\end{figure}

ED calculations were carried out on $2\times L$ ladders
with $L$ an odd number of sites.  Both periodic and anti-periodic boundary
conditions for $L$ up to 13 were used \cite{note2}.  In Fig.~2 we show results for the
triplet excitation energies in the ${\mathbf K}=(\pi,\pi)$ sector 
for a sequence of
$2\times L$ ladders with 2 holes for various values of $t^\prime$. Here the
excitation energy is measured relative to the 2-hole ${\mathbf K}=(0,0)$ 
ground
state.  The lowest triplet ${\mathbf K}=(\pi,\pi)$ state is found to be 
separated from
a quasi-continuum of higher energy states. In Fig.~2, the error bars mark
the difference between the results obtained using periodic and anti-periodic
boundary conditions with the open symbols marking the mean value. Since the
actual longitudinal momentum for a finite ladder is $\pi(1- 1/L)$, we have
extrapolated these results using a scaling form $A+B/L + c/L^2$. The solid
symbols denote the DMRG calculation of the spin gap $\Delta_S$,
Eq.~(\ref{two}), for an open $2\times 32$ ladder with 2 holes
(larger lattices are also included for $t'=0$).  For $-0.2
\leq t^\prime \leq 0.5$, the extrapolated ED results for the 
${\mathbf K}=(\pi,\pi)$
triplet pass through the DMRG spin gap $\Delta_S$. However, for
$t^\prime=-0.5$, the DMRG determined spin gap lays well below the
extrapolated ${\mathbf K}=(\pi,\pi)$ triplet.  
As discussed in the introduction, for
$-0.2 \leq t^\prime$, the spin gap is set by the excitation in the triplet
${\mathbf K}=(\pi,\pi)$ sector.  
However, for $t^\prime \ltwid -0.2$, the spin gap is
set by the onset of the two quasi-particle continuum $\Delta_P$
which goes to zero as $t^\prime$ approaches
$-0.5$. 
Note that, even when $\Delta_S < \Delta_{MP}$, the magnon-hole pair state
could still be {\it locally} stable if the decay process into  
2 quasi-particles with the same momentum $(\pi,\pi)$ 
is impossible.
Although the ED results approach the 
${\mathbf K}=(\pi,\pi)$ magnon energy of
the undoped ladder for $t^\prime > 0.35$, DMRG results show that the 
character of
the triplet excitation changes from a bound magnon-hole-pair to a separate
magnon and hole-pair state.  Thus, in this regime, the spin gap is set by
the excitation energy of the magnon $\Delta_M$ and is therefore
continuous upon doping.
\begin{figure}
\begin{center}
\psfig{figure=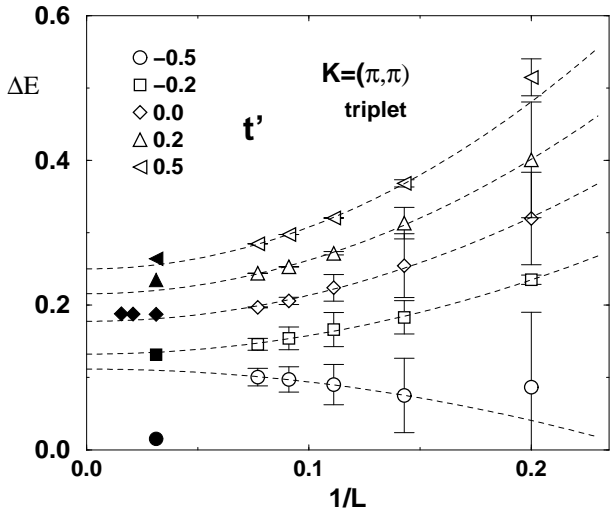,width=8truecm,angle=-90}
\end{center}
\vspace{-0.2truecm}
\caption{A finite-size scaling analysis of the lowest triplet eigenenergies
of a $2\times L$ ladder with 2 holes in the 
${\mathbf K}=(\pi,\pi)$ sector measured
with respect to the ${\mathbf K}=(0,0)$ ground state singlet for various
values of $t^\prime$.  
The open symbols are ED
data obtained on closed ladders up to $2\times 13$ in size. The mean value
between periodic and antiperiodic boundary conditions is denoted by the
open symbols. The error
bars are given by the energy difference between the two boundary conditions.  
The full
symbols are DMRG results for the spin gap $\Delta_S$ defined by
Eq.~(\ref{two}).
}
\label{extra}
\end{figure}

In order to get a clearer picture of the nature of the triplet excitations
which determine the spin gap, we have used DMRG results to study the spin
and hole correlations in these states. In Fig.~3 a center section of
a $2\times 32$ ladder with $t^\prime=0$ is shown.  The upper and middle
diagrams show the probability of finding the second hole when the first
hole is projected out at the center of the upper leg for the singlet
\cite{WS97} and
triplet state respectively.  In both of these states, the two holes are
bound and the most probable configuration for $J/t=0.5$ corresponds to
having the holes on diagonal sites. Note that the triplet-bound state is
more extended than the singlet bound state.  The lowest diagram shows the
spin distribution for the $S_z=1$ triplet state when the two holes are
projected out at their most probable sites.  
It is clear that this state is a bound magnon-hole pair.

\begin{figure}
\begin{center}
\psfig{figure=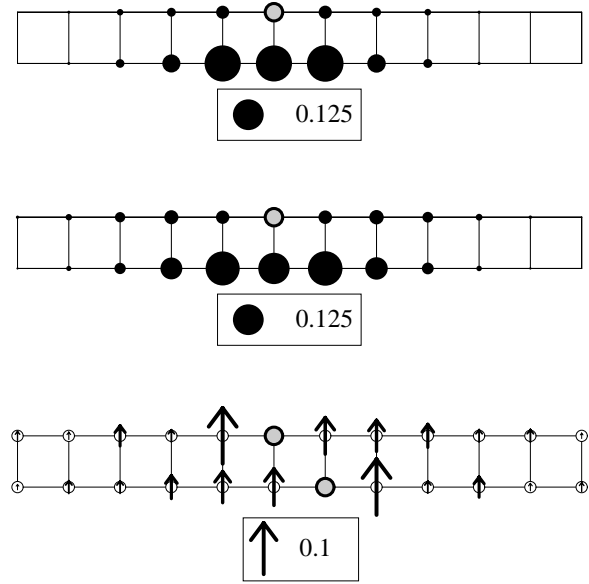,width=8truecm,angle=0}
\end{center}
\vspace{-0.2truecm}
\caption{Structure of the ground state and the lowest energy 
triplet state 
of a 2-leg ladder with 2 holes and $t^\prime=0$. In the top and middle
diagrams, the diameter of the black circles is proportional to the
probability of finding the second hole when one hole is projected out at
the center of the upper leg for the ground singlet
and lowest energy triplet states
respectively.  Note that the triplet state is more extended than the
singlet bound state. 
The bottom diagram provides the spin distribution in the
triplet state.
}
\label{fig3}
\end{figure}

Similar calculations for $t^\prime=-0.5$ show that in the triplet state the
two holes are unbound while for $t^\prime=0.5$ the two holes are bound into
a singlet and 
uncorrelated with the spin 1 excitation.  This behavior is shown in Fig.~4,
where we have plotted
\begin{equation}
\left\langle S_z(\ell_x)\right\rangle \equiv \left\langle S_z(\ell_x, 1)
P_h(i)P_h(j)\right\rangle/\left\langle P_h(i) P_h(j)\right\rangle
\label{five}
\end{equation}
versus $\ell_x$ for $t^\prime=0$, $-0.5$, and 0.5.  Here $P_h(i)$ is the
projection operator for a hole at the $i^{\rm th}$ site. For $t^\prime=0$, we
have set $i=(16,2)$ and $j=(17,1)$, corresponding to the most probable hole
location. Here, as previously illustrated in Fig.~3, we see that
the spin is bound to
the hole-pair.  For $t^\prime=0.5$ we again have a situation where 
the holes are most likely to sit close to each other, 
and here we have projected them onto $i=(16,1)$ and $j=(16,2)$.  
However, in this case, the spin 1 is
spread out corresponding to a magnon which is not bound to the hole-pair.
Finally, for $t^\prime = -0.5$, one finds that the lowest energy triplet
excitation corresponds to two separate quasi-particles.

We finish with a brief discussion of some experimental
results for the superconducting two-leg doped 
ladder Sr$_2$Ca$_{12}$Cu$_{24}$O$_{41}$
\cite{Jer98}. Nuclear magnetic resonance measurements of the copper-63
Knight shift and relaxation time $T_1$ suggest a collapse of the spin
gap with pressure.
We believe this signals the appearance of new low lying triplet 
excitations upon doping the ladder planes and points towards a
negative value of $t'$ \cite{cuprates}. 
In this regime, due to the presence of a low-energy quasi-particle continum
located predominantly around the zone center,
momentum-resolved experiments like inelastic neutron scattering 
would be essential to search for sharp finite energy triplet excitations.

\begin{figure}
\begin{center}
\psfig{figure=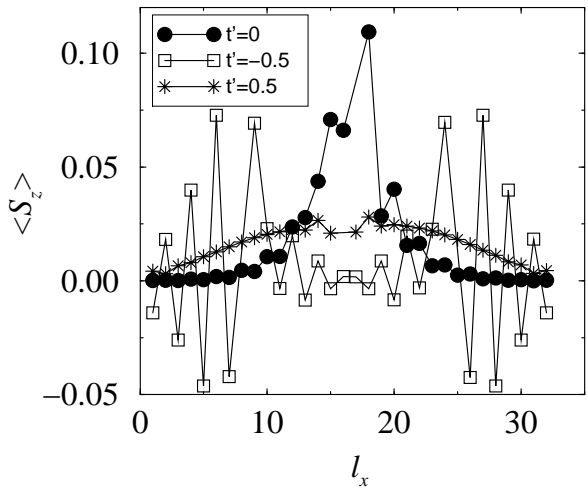,width=8truecm,angle=0}
\end{center}
\vspace{-0.2truecm}
\caption{Here, for three different values of $t^\prime$ we show
$\left\langle S_x(\ell_x)\right\rangle$ (as defined by Eq.~(\ref{five})) 
versus $\ell_x$. For $t^\prime=0$, we take $i=(16,2)$, and $j=(17,1)$, one 
of the most probable locations for the two holes, and one sees that a magnon 
is tightly bound to the hole pair.  For $t^\prime=0.5$, we have $i=(16,1)$ and
$j=(16,2)$, also one of the most probable locations for the holes. Here we
see a magnon which is not bound to the hole pair.  For $t^\prime=-0.5$, in
order to illustrate how separate $S=1/2$ spins are bound to each hole, we
plot $\langle S_z(\ell_x)\rangle$ with $i=(8,1)$ and $j=(25,1)$.
}
\label{fig4}
\end{figure}

To summarize, using ED and DMRG calculations, we have
found that the spin gap can evolve in different ways when two holes are
doped into  a 2-leg
ladder.  When the 2 holes are added, it is possible that the lowest energy
triplet state simply remains the 
${\mathbf K}=(\pi,\pi)$ magnon so that there is no
change in $\Delta_S$. In this case the 2 added holes remain in a bound
$d_{x^2-y^2}$-like singlet state and a triplet magnon similar to that of an
undoped ladder is created.  As the length of the ladder increases, the
interaction between these two entities becomes negligible. We see this
happening for $t^\prime\gtwid 0.35$. It is also possible that the lowest
energy triplet state has 
${\mathbf K}=(0,0)$ and is set by the two-quasi-particle
continuum corresponding to the pair-binding energy $\Delta_P$. In this
case, there is a discontinuous change in the spin gap upon doping and the
lowest energy triplet state arises from the dissociation of a pair into two
quasi-particles. We see this
for the present model when $t^\prime \ltwid - 0.2$. Finally, for the
intermediate region $-0.2 \ltwid t^\prime \ltwid 0.35$ we find that the
lowest energy triplet state has ${\mathbf K}=(\pi,\pi)$ 
and corresponds to a bound
magnon-hole pair with energy $\Delta_{MP} < \Delta_M$. In this case, there
is again a discontinuous evolution in the spin gap from $\Delta_M$ to
$\Delta_{MP}$ upon doping. 

We would like to acknowledge useful discussions with Ian Affleck. 
S.R.~White and D.J.~Scalapino acknowledge support from the NSF under
grant \# DMR98-70930 and grant \# DMR98-17242 respectively.
D.~Poilblanc thanks IDRIS (Paris) for allocation of CPU time on the NEC SX5
supercomputers.

\end{document}